# Level Set Method for Quantum Control of Dipole Moment

**Fariel Shafee**
Department of Physics
Princeton University
Princeton, NJ 08540

*Abstract:*

We investigate the level set method (LSM) in a specific quantum context; namely the dipole transition moment for a system with a nontrivial Morse potential. We draw equal moment sets in the two-dimensional space of two important parameters of the potential, namely the depth of the potential and its width. Another variable is introduced as a scale and we see "motions" of the level sets normal to the contours, as in classical contexts such as fluid dynamics or in epitaxial crystal growth. Presumably interpolating the level sets normally by smooth functions such as splines may give a fairly accurate method of combining the variables to keep the dipole moment invariant.

## I. Introduction

The expectation value of any observable is a functional of the parameters of the Hamiltonian and its calculation in analytic terms in a compact closed form is usually possible only in trivial cases. On the other hand numerical methods are also highly expensive with respect to computer memory and time and for quantum control in real time of fast molecular and atomic processes we need methods for simplifying the algorithm without sacrificing much accuracy.

Geremia and Rabitz [1,2] have shown that the high dimensional model representation (HDMR) is a quite effective method for forward mapping of the parameters of the Hamiltonian and the functional form of the potential as low order finite sums of the component correlations can in many cases faithfully reproduce random test situations after adjustments to minimize a cost function that takes care of both fit to the data as well as of expected rigidity and smoothness of the resultant form.

In a previous work [3] we have shown that the energy levels of a molecule in harmonic motion with a background perturbing time independent field can be fitted smoothly to "data" points by a B-spline surface. It retains a minimal curvature avoiding wild excursions natural for high degree multinomials forced to go through a large number of points, because it is actually a function of a lower degree which is infinitely differentiable at all points except a finite number of points where it has only the lower derivative(s) continuous.

In the HDMR method also one ignores the higher order correlations between the functional variables, in a fashion reminiscent of perturbation theory where multiple correlations are usually taken as insignificant.

In the LSM the picture is even simpler as the variation of the object of interest in terms of a scale parameter standing in for the time variable present in the case of real motion cases is again described by pseudo-velocities and gradients of the contours representing the level sets. Since the gradient is the first derivative, continuity of the gradient needed in this picture, and ensured usually by forcing in an anti-curvature cost term, effectively reduces the problem so a lower degree quasi-local problem, which looks smooth like an analytic formula.

## II. The Potential and the Parameters

We shall use the Morse potential in this work, though the methods are of course equally applicable in a wider context and we have chosen the ground to first excited state of transition dipole moment as the observable.

$$V(r) = C\,[\exp\{-2a(r-r_0)\} - 2\exp\{-a(r-r_0)\}] \tag{1}$$

It is well-known that such potentials have only a finite number of bound states depending on the depth $C$ and the width $1/a$ of the potential. However we shall be concerned with only the lowest order transition moment, so it is sufficient to have only two states. Numerical solution shows that for $V > 6$, and $a \sim 1$, we always get at least two solutions.

In this work we are not trying to reproduce any specific molecular values, but typically we expect $r_0$ and $1/a$ to be of the order of an angstrom, and $C$ to be of the order of meV.

The transition moment is of course

$$d_{01}(a(s), C(s)) = <0|x|1>  \quad (2)$$

where the scale (*s*) dependence of *a* and *C* can be complicated or simple as the case may be. The dipole moment has no direct *s* dependence, but acquired it implicitly through *a* and *C* after the integration over the wave functions |0> and |1>.

Let us assume for our purpose a simple *s* dependence for *a* and *C*:
$$a \rightarrow a/s$$
$$C \rightarrow Cs \quad (3)$$
where we have used the fact that *C* is energy and hence has the inverse dimension as *1/a*.

So a constant $d_{12} = c$ surface has the "dynamical" equation

$$u . \nabla d_{01} = 0 \quad (4)$$

where

$$u_{a(C)} = da(C)/ds \quad (5)$$

i.e. the "velocity" of the parameters is orthogonal to the equal $d_{12}$ level set.

### III.  Numerical Results

In Fig. 1 we show the level sets for *s= 1, 1.1, 1.2, 1.5, 1,7* and *1.8* with *C* from *12* to *26* and *a* from *0.8* to *2*.

The strange kinks for the highest *s* after *C = 21* are numerical hiccups whose exact cause has not yet been located.

There are two notable features of the level sets.
1. The are more or less parallel over the region in the figure and hence the normal to one will be nearly normal to the other contours too.
2. However, the speed too does not vary much. Hence information from local data is not so important get the speed at each level (i.e. at each *s*). What we are witnessing here is kind of global "force" and "acceleration" on the whole contour, and small variations can probably be taken care of by an interpolation procedure such as a spline fit or the HDMR method.

The average speeds for each s value collected from a number of points on the corresponding levels is shown in Table I.

## Table I

| s | 1.0 | 1.1 | 1.2 | 1.5 | 1.7 |
|---|---|---|---|---|---|
| \|du/ds\| | 1.35 | 1.38 | 1.36 | 1.29 | 1.29 |

One can make a crude approximation of the advancement of the level sets in s using such a gross velocity since the speed is nearly constant. The curvature at low *C* will also propagate with parallel displacement.

## IV.  Conclusions

The LSM ideas as presented here can definitely be made more refined. The best strategy may be a hybrid of splines, HDMR and LSM, to get the optimal accuracy and cost. We have not considered the question of errors here at all. Inclusion of errors will make the level sets effectively annuli and not curves. This opens up the possibility of using HDMR type handling of best fits with curvature damping costs included. With many degrees of freedom allowed by the fuzzy curve, it may be necessary to use Genetic Algorithm to minimize the cost function. This is now being investigated. It is also necessary to check whether other types of observables such as scattering cross-sections, or control problems involving laser pulse fields are also amenable to such an approach.

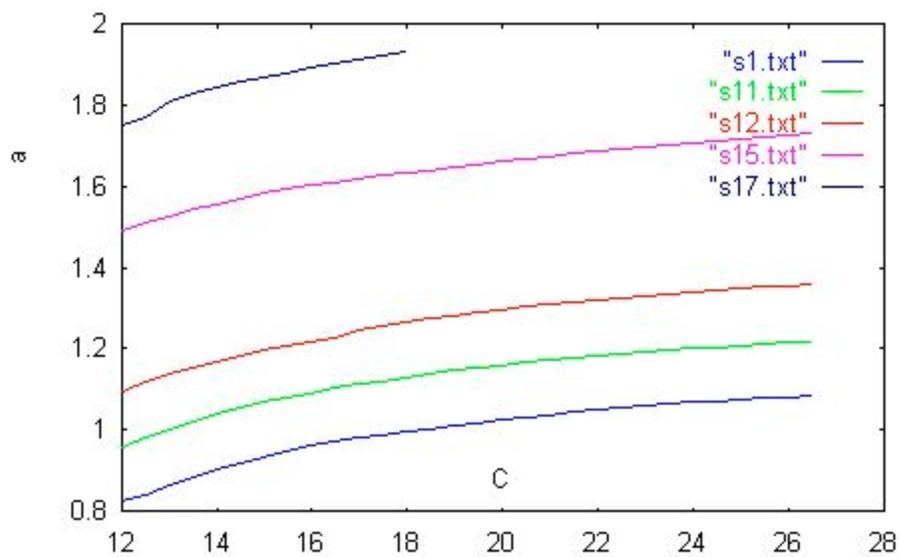

Fig. 1: Level sets for the lowest transition dipole moment for a Morse potential with depth *C* and width *a* at scale values of *s=1,1.1, 1.2, 1.5* and *1.7*